\documentclass[10pt,twocolumn,letterpaper]{article}

\usepackage{cvpr}              

\usepackage{graphicx}
\usepackage{amsmath}
\usepackage{amssymb}
\usepackage{booktabs}
\usepackage{multirow}

\usepackage[pagebackref,breaklinks,colorlinks]{hyperref}

\usepackage[capitalize]{cleveref}
\crefname{section}{Sec.}{Secs.}
\Crefname{section}{Section}{Sections}
\Crefname{table}{Table}{Tables}
\crefname{table}{Tab.}{Tabs.}

\def\cvprPaperID{*****} 
\def\confName{CVPR}
\def\confYear{2023}

\begin{document}

\title{AudioInceptionNeXt: TCL AI LAB Submission to EPIC-SOUND Audio-Based-Interaction-Recognition Challenge 2023}

\author{Kin Wai Lau, \space Yasar Abbas Ur Rehman, \space Yuyang Xie, \space Lan Ma \\
TCL AI Lab\\
 {\tt\small \{stevenlau, yasar, yuyang.xie, rubyma\} @tcl.com } 
}
\maketitle

\begin{abstract}
   This report presents the technical details of our submission to the 2023 Epic-Kitchen EPIC-SOUNDS Audio-Based Interaction Recognition Challenge. The task is to learn the mapping from audio samples to their corresponding action labels. To achieve this goal, we propose a simple yet effective single-stream CNN-based architecture called AudioInceptionNeXt that operates on the time-frequency log-mel-spectrogram of the audio samples. Motivated by the design of the InceptionNeXt, we propose parallel multi-scale depthwise separable convolutional kernels in the AudioInceptionNeXt block, which enable the model to learn the time and frequency information more effectively. The large-scale separable kernels capture the long duration of activities and the global frequency semantic information, while the small-scale separable kernels capture the short duration of activities and local details of frequency information. Our approach achieved 55.43\% of top-1 accuracy on the challenge test set, ranked as $1^{st}$ on the public leaderboard. Codes are available anonymously at \url{https://github.com/StevenLauHKHK/AudioInceptionNeXt.git}.
\end{abstract}

\section{Introduction}
\label{sec:intro}
Learning feature representations for audio event classification has been extensively studied over the past decade using a variety of deep neural network architectures like Convolutional Neural Networks (CNN), Long-Short Term Memory (LSTM), and the recent state-of-the-art Transformer networks. These models have achieved remarkable performance on a number of audio-based event classification datasets, such as AudioSet \cite{gemmeke2017audio}, VGGSound \cite{chen2020vggsound}, EPIC-KITCHENS-100 \cite{damen2020rescaling} and the recent EPIC-SOUNDS \cite{EPICSOUNDS2023}. The former two datasets are collected from YouTube that contain a variety of event activities, such as classifying different musical instruments, animal sounds, vehicle sounds and home activities sounds. The latter two datasets are captured from egocentric videos, which contain unscripted daily activities and the interactions among different objects in the kitchen. EPIC-SOUNDS is a recently proposed kitchen event classification dataset, derived from the audio of EPIC-KITCHENS-100. This new dataset tackles two annotation issues in the EPIC-KITCHENS-100, i.e., temporal misalignment between visual and auditory events, and a single class label used for both visual and audio modalities. Meanwhile, this dataset introduces several challenges, for example, variable lengths of audio associated with different activities, and background sound captured with the event activities. 

To solve the aforementioned challenges, the recent approaches train Transformers either using supervised learning \cite{gong2021ast, chen2022hts} or self-supervised learning \cite{gong2022ssast, baevski2020wav2vec}. 
While Transformers-based architecture is the most commonly used method nowadays, we focus on a relatively underexplored domain of CNN-based architectures. Our motivation for investigating the CNN-based architecture is due to two reasons. First, these architectures are still prevalent for audio classification tasks due to their low computational cost and memory footprint. Second, their performance is either comparable to or better than the state-of-the-art Transformers models \cite{wang2022towards}. For instance, \cite{kazakos2021slow} proposed a two-stream CNN-based network, called the Slow-Fast model for learning audio representations. The Slow stream takes a lower temporal resolution of the audio spectrogram input, focusing on the global frequency semantic information and long-term activities. In contrast, the Fast stream adopts the full high-resolution input, focusing on the local frequency information and short-term activities. Later work \cite{wang2022towards} extends the study of the Slow-Fast model with self-supervised contrastive learning and demonstrated that this model performs better than the state-of-the-art ViT transformer model \cite{gong2021ast} on a number of downstream tasks. 

\begin{figure*}[h]
\centering
\includegraphics[width=\linewidth]{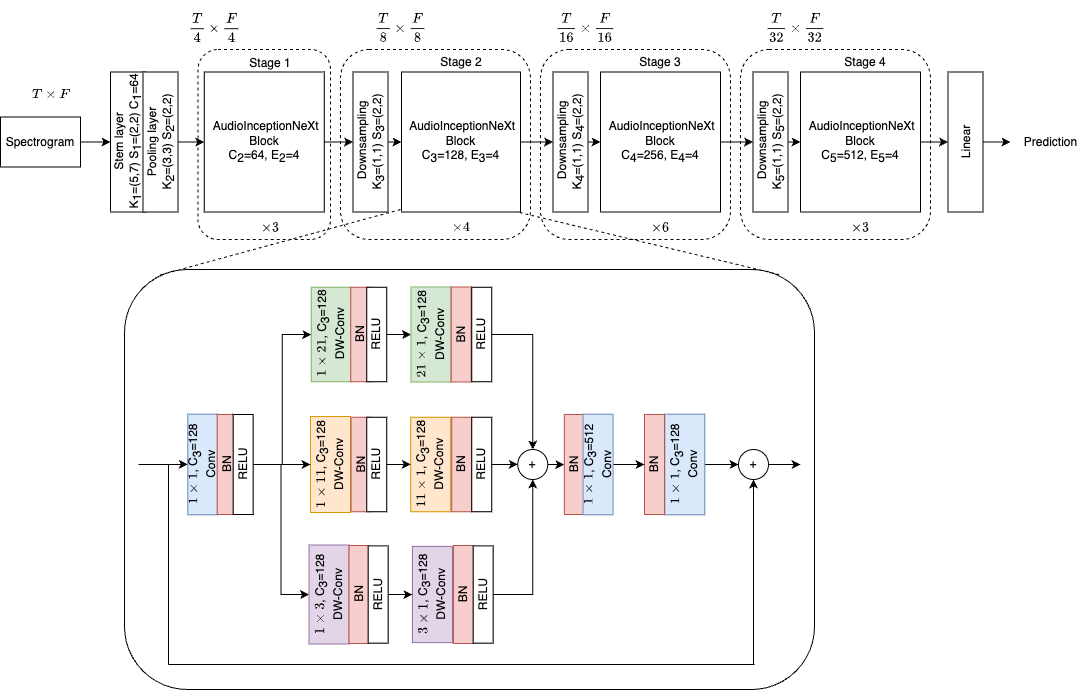}
\caption{Proposed AudioInceptionNeXt architecture. The top of the figure shows the macro design of the AudioInceptionNeXt, and the bottom one shows the micro block design.}
\label{fig:architecture}
\end{figure*}

In this work, we propose a simple yet effective single-stream CNN model for audio event classification. Specifically, we re-design the conventional CNN residual block and employ multi-scale separable convolutional kernels to capture the global and local time-frequency information effectively. Following the success of the large kernel in CNN-based architectures \cite{ding2022scaling, guo2022visual, liu2022convnet}, we use the kernel size of 3, 11, and 21 in AudioInceptionNeXt block. The large kernel captures the global frequency semantic information and long-term activities, while the small kernel captures the local details of frequency information and short-term activities. Experiments demonstrate that our model outperforms previous CNN-based models and transformer-based ViT models on the EPIC-SOUNDS validation set.
\section{Methodology}
\label{sec:method}
In this section, we first describe the macro architecture design of the proposed AudioInceptionNeXt, followed by the micro block design.
\subsection{Macro Architecture Design}
 We follow the hierarchical design of Resnet50 \cite{he2016deep} as shown in Fig.\ref{fig:architecture}. The hyperparameters of the model are listed below.
\begin{itemize}
\item {${S_i}$: the stride of the convolution layer in the input stem and downsampling in stage $i$;}
\item {${K_i}$: the kernel size of the convolution layer in input stem and downsampling in stage $i$;}
\item {${C_i}$: the number of output channels in stage $i$;}
\item {${E_i}$: the channel expansion ratio of inverted bottleneck in stage $i$;}
\end{itemize}

\textbf{Model Input.} As shown in Fig.\ref{fig:architecture}, the AudioIncepionNeXt is fed with the log-mel-spectrogram of the audio signal. The log-mel-spectrogram has a resolution of  $T \times F$, where the T and F axes represent the time and frequency bin, respectively.

\textbf{Macro Design.} Adhering to the design principles of ResNet50, our model comprises an input stem layer followed by four subsequent stages. The stem layer consists of a $5 \times 7$ convolutional layer with a stride of 2 and outputs 64 channels. It is followed by a $3 \times 3$ max-pooling layer with a stride of 2 for subsequent downsampling of the spatial resolution (time and frequency axes) of the convolutional feature maps. Except in stage 1 following the stem layer, the AudioInceptionNeXt block in each stage is preceded with a $1 \times 1$ convolutional layer with a stride of 2 for downsampling the spatial resolution of the convolutional feature maps.

The AudioInceptionNeXt block contains $1 \times 1$ convolutional layers and multi-branch convolutional layers with multi-scale depthwise separable kernels. We adopt the kernel size of 3, 11, and 21 for the multi-scale depthwise separable kernels in our experiments. Like ConvNeXt \cite{liu2022convnet}, we adopt the inverted bottleneck design in each AudioInceptionNeXt block, i.e., the channel size of the hidden $1 \times 1$ convolutional layer preceding the last $1 \times 1$ convolutional layer is four times wider than the input along the channel dimensions, as shown in Figure \ref{fig:architecture}.

\subsection{Micro Architecture Design}
\textbf{Parallel multi-scale kernel.} To capture the multi-scale temporal and frequency information using a single-stream network, we adopt the multi-branch design of the visual CNN-based InceptionNet \cite{szegedy2016rethinking}. The core component of the InceptionNet is the multi-branch convolutional layers which contain multi-scale kernel sizes for capturing the different scales of the objects. In our proposed block, we follow a similar design by using small and large kernels in different branches. The large kernel (i.e., $1 \times 11$ and $11 \times 1$, $1 \times 21$ and $21 \times 1$) captures the global frequency semantic information and long-term activities. In contrast, the small kernel (i.e., $1 \times 3$ and $3 \times 1$) captures the local details of the frequency information and short-term activities. Finally, all the feature maps from different branches are added and passed to the $1 \times 1$ convolutional layers for channel-wise information exchange.

\textbf{Depthwise separable kernel.} Using a large kernel in the convolutional layer is both computationally and memory inefficient. Following kernel design in ConvNeXt and InceptionNeXt \cite{yu2023inceptionnext}, we adopt the depthwise convolution layer with the separable kernel. In this design, the kernel size of  $x \times y$ is decomposed into $1 \times x$ and $y \times 1$ kernels. We show such decomposition for $21 \times 21, 11 \times 11$,  and $3 \times 3$ kernel in the in Fig.\ref{fig:architecture}. Apart from saving the computational time and memory footprint, \cite{kazakos2021slow, xiao2020audiovisual} has shown that a separable kernel allows the model to extract the temporal and frequency feature independently, which improves the audio classification results. The reason is that the statistics of the spectrogram are not homogeneous, unlike the natural images.

\textbf{Inverted bottleneck.} Unlike the conventional design of inverted bottleneck in MobileNetV2 \cite{sandler2018mobilenetv2}, we place the multi-branch convolutional layers at the top before applying the $1 \times 1$ convolutional layers for conducting the channel-wise expansion and squeeze operation, similar to ConvNeXt block \cite{liu2022convnet}. This helps to save the memory footprint and computational time caused by the large kernel design.

\textbf{Non-linear layers.} To increase the non-linearity in model, each separable kernel is followed by batch normalization and RELU activation layers as shown in the magnified view of the InceptionNeXt block in Fig.\ref{fig:architecture}.

\begin{table}[h]
        \caption{Comparsion with CNN-Based SOTA methods on pre-train VGG-Sound event classification dataset. The results are reported on the VGG-Sound validation set. Param represents the parameter size of the model. GFLOPs represent the floating point operation. Top-1 and Top-5 represent the top-1 accuracy and top-5 accuracy, respectively. mAP and AUC represent the mean average precision and area under the curve, respectively.}
        \resizebox{1\columnwidth}{!}{
        \begin{tabular}{l|c|c|c|c|c|c|c}
            \hline
             \textbf{Model} & \textbf{Param} & \textbf{GFLOPs} & \textbf{Top-1} & \textbf{Top-5} & \textbf{mAP} & \textbf{AUC} & \textbf{d-prime} \\
            \hline
            ResNet50 \cite{jansen2018unsupervised} & 24.13 & 5.26 & 52.07 & 77.72 & 54.1 & 97.3 & 2.74 \\
            InceptionNeXt Tiny \cite{yu2023inceptionnext} & 24.20 & 5.46 & 50.16 & 76.28 & 52.5 & 97.4 & 2.75  \\
            Slow-Fast \cite{kazakos2021slow} & 26.68 & 5.55 & \textbf{52.24} & \textbf{78.14} & \textbf{54.4} & 97.5 & 2.76 \\
            AudioInceptionNeXt (ours) & \textbf{11.83} & \textbf{2.62} & 51.94 & 77.95 & 54.1 & \textbf{97.6} & \textbf{2.79} \\
            \hline
        \end{tabular}
        }
    \label{tab:pretrain-baseline-comparsion}
\end{table}

\begin{table*}[h]
    \centering
    \caption{Transfer learning results on EPIC-SOUNDS validation set. mPCA represents the mean per-class accuracy. mAUC represents the mean area under the curve.}
    \resizebox{1.8\columnwidth}{!}{
    \begin{tabular}{l|l|ccccccc}
        \hline
        Type & Model & & \multicolumn{4}{c}{EPIC-SOUNDS} \\
        & & Params.(M) &	GFLOPs & Top-1 & Top-5 & mPCA & mAP & mAUC \\
        \hline
         \multirow{4}{*}{CNN-based} & ResNet50\cite{jansen2018unsupervised} & 23.59 & 4.27 & 52.57 & 82.77 & \textbf{21.21} & 0.238 & 0.864 \\
         & InceptionNeXt Tiny\cite{yu2023inceptionnext} & 24.00 & 4.43 & 51.24 & 81.73 & 20.68 & 0.227 & 0.865 \\
        & Slow-Fast (baseline)\cite{kazakos2021slow} & 26.06 & 4.50 & 52.84 & 83.12 & 20.74 & 0.242 & 0.860 \\
        & AudioInceptionNeXt (ours) & \textbf{11.69} & \textbf{2.13} & \textbf{54.05} & \textbf{84.54} & 20.91 & \textbf{0.244} & \textbf{0.875}\\
        \hline
        Transformer & SSAST \cite{gong2022ssast} & 87.22 & 48.67 & 53.47 & 84.37 & 20.22 & 0.234 & 0.842\\
        \hline
        \end{tabular}
        }
    \label{tab:transfer-epic-sound}
\end{table*}

\section{Experiments}
\subsection{Training and Validation Details}
We follow the same training strategy as the baseline Slow-Fast model \cite{kazakos2021slow}. We first pre-train our model on the VGG-Sound dataset \cite{chen2020vggsound} and then fine-tune it on the EPIC-SOUNDS dataset \cite{EPICSOUNDS2023}.  We use the Librosa library to convert the audio raw audio signal to log-mel-spectrogram with 128 mel bands before feeding it into the network. As a common practice in \cite{kazakos2021slow, gong2021ast, gong2022ssast, chen2022hts}, we apply SpecAugment \cite{park2019specaugment} argumentation method in both training stages (pertaining and fine-tuning) that contain frequency masking, time masking, and time warping. We trained all our models with batch size 32 on 4x NVIDIA RTX3090 GPUs in pretraining and fine-tuning.

\subsubsection{Pretraining}
 In the pretraining stage, we randomly pick a 5.12 seconds audio and apply the log-mel filterbanks with a window size of 20ms, and a hop length of 10ms. This results in a spectrogram of size $512 \times 128$ which is fed to the model as an input. We randomly initialize the model and optimize it using SGD for 50 epochs with a momentum of 0.9 and a learning rate of 0.01. We drop the learning rate by 0.1 at epochs 30 and 40.
 
\subsubsection{Fine-Tuning}
In the fine-tuning stage, we randomly pick 2.08 seconds of audio and apply log-mel filterbanks with a window size of 10ms and a hop length of 5ms. This results in a spectrogram of size $416 \times 128$. Note that edge padding is applied to the spectrogram if the resulting size is lower than the predefined resolution. We attach a linear prediction head on top of the VGG-Sound pre-trained backbone model to classify the 44 action classes in the EPIC-SOUNDS dataset \cite{EPICSOUNDS2023}. We freeze all the batch normalization layers except the first one in the stem layer and fine-tuned the whole model. We use the same optimizer setting as the pretraining stage except the initial learning rate is set to 0.001, which is reduced after 20 and 25 epochs by a factor of 0.1. The model is finetuned for 30 epochs. 

\subsection{Results}
\noindent\textbf{Results on VGG-Sounds:} We validate the effectiveness of the proposed AudioInceptionNeXt model against SOTA CNN-based models using the VGG-Sound validation set. Specifically, the comparison is performed against a series of CNN-based models that include ResNet50\cite{jansen2018unsupervised}, InceptionNeXt Tiny\cite{yu2023inceptionnext} and Slow-Fast model \cite{kazakos2021slow}. The results are reported in Table \ref{tab:pretrain-baseline-comparsion}. We note the following observations: (1) The AudioInceptionNeXt model can save more than half of the parameters and GFLOPs while obtaining only a minor performance drop of 0.3\% and 0.13\% compared to the baseline Slow-Fast model and ResNet50, respectively. (2) The AudioInceptionNeXt model  outperforms the similar multi-branch design InceptionNeXt network by 1.78\% and 1.67\% in terms of top-1 and top-5 accuracy respectively, while saving more than 50\% of parameters and GFLOPs.\\ 

\noindent\textbf{Results on EPIC-SOUNDS:}
We conducted a comprehensive comparison between our proposed model, other SOTA CNN-based models, and the Transformer-based model on the EPIC-SOUNDS dataset. As shown in Table \ref{tab:transfer-epic-sound}, our model outperforms the Slow-Fast model, ResNet50, and InceptionNeXt by 1.21\%, 1.48\%, and 2.81\%, respectively in Top-1 accuracy. The AudioInceptionNeXt requires 50\% fewer parameters and incurs 50\% fewer GLOPs compared to the CNN-based models. Interestingly, the AudioInceptionNeXt performed much better than the Transformer-based SSAST model while incurring only 11.69M (vs. 87.22M) parameters and 2.13 GFLOPs (vs. 48.67 GLOPs). This results in approx. 86\% parameter saving and a reduction in 95\% GLOPs.
\section{Conclusion}
In this technical report, we present a simple and effective single-stream CNN-based model called AudioInceptionNeXt. This model employs a multi-branch design like InceptionNet with multi-scale separable depth-wise convolutional kernels. Our experiments demonstrate that the proposed model achieves a SOTA performance in the EPIC-SOUNDS dataset compared to the previous CNN-based models and the Transformer-based model while obtaining the lowest computational GFLOPs and parameter size. 

{\small
\bibliographystyle{ieee_fullname}
\bibliography{egbib}
}

\end{document}